\documentclass[twocolumn]{jpsj2} 
\def\gsim{\,$\raise0.3ex\hbox{$>$}\llap{\lower0.8ex\hbox{$\sim$}}$\,}
\def\lsim{\,$\raise0.3ex\hbox{$<$}\llap{\lower0.8ex\hbox{$\sim$}}$\,}

\title{
Phase transitions in spin-orbital coupled model for pyroxene titanium oxides
}

\author{
Toshiya \textsc{Hikihara}$^{1}$\thanks{E-mail address: hikihara@phys.sci.hokudai.ac.jp} 
and  
Yukitoshi \textsc{Motome}$^{2}$
}

\inst{
$^{1}$Division of Physics, Hokkaido University, Sapporo 060-0810, Japan\\
$^{2}$RIKEN (The Institute of Physical and Chemical Research), 
Wako, Saitama 351-0198, Japan \\
}

\abst{
We study the competing phases and the phase transition phenomena 
in an effective spin-orbital coupled model derived for 
pyroxene titanium oxides $A$TiSi$_2$O$_6$ ($A$=Na, Li).
Using the mean-field-type analysis and 
the numerical quantum transfer matrix method, 
we show that the model exhibits two different ordered states, 
the spin-dimer and orbital-ferro state and 
the spin-ferro and orbital-antiferro state. 
The transition between two phases is driven by 
the relative strength of the Hund's-rule coupling 
to the onsite Coulomb repulsion 
and/or by the external magnetic field. 
The ground-state phase diagram is determined.
There is a keen competition between orbital and spin degrees of freedom 
in the multicritical regime, 
which causes large fluctuations and significantly affects 
finite-temperature properties in the paramagnetic phase. 
}

\kword{pyroxene titanium oxides, orbital degree of freedom, 
Kugel-Khomskii model, spin singlet, dimerization, 
Hund's-rule coupling, superexchange interaction, 
field-induced transition, quantum transfer matrix method}

\begin{document}
\maketitle

\section{Introduction} 

It has been recognized 
that the orbital degree of freedom plays an important role 
in transition metal compounds.\cite{Kugel1982,Tokura2000,Imada1998}
The interplay among the orbital, spin, and charge degrees of freedom 
produces a rich variety of phenomena.
In particular, quantum and thermal fluctuations in the competition 
between the spin and orbital exchange interactions 
have attracted much interest. 

Pyroxene titanium oxides $A$TiSi$_2$O$_6$ ($A$=Na, Li) are typical examples 
of the systems where the interplay between orbital and spin 
is expected.\cite{Isobe2002}
In these insulating materials, 
each magnetic Ti$^{3+}$ cation has one $d$ electron in $t_{2g}$ levels.
The lattice structure consists of 
characteristic one-dimensional (1D) chains of 
skew edge-sharing TiO$_6$ octahedra, which 
are bridged and well separated by SiO$_4$ tetrahedra.
Hence, we can regard the system as a quasi 1D spin-$1/2$ system.

These pyroxene titanium oxides 
exhibit a phase transition to the low-temperature phase 
with a spin gap and a lattice dimerization 
at $T \simeq 200$K,\cite{Isobe2002,Ninomiya2003}
which reminds us of a spin-Peierls transition.
However, the magnetic susceptibility shows a peculiar temperature dependence 
clearly distinct from that of usual spin-Peierls compounds.\cite{Isobe2002} 
The magnetoelastic effect is not sufficient to explain the data.
Therefore, the other scenario is needed 
to capture the physics of the transition.

In the previous study, we derived an effective spin-orbital 
coupled model for the pyroxene compounds in the strong correlation limit 
of a multiorbital Hubbard model, 
and investigated it using the numerical method 
as well as mean-field (MF) analyses.\cite{Hikihara2004}
We have shown that the interplay between orbital and spin degrees 
of freedom plays a decisive role in the spin-singlet formation. 
In particular, we have found that the system exhibits 
a nontrivial feedback effect between the orbital and spin correlations: 
At high temperatures, 
both correlations are antiferro (AF) type and compete with each other.
When the Hund's-rule coupling is small enough 
compared to the onsite Coulomb repulsion, 
the AF spin correlation 
wins the competition and grows rapidly 
as temperature decreases.
The enhancement of the spin correlation changes the orbital correlation 
from AF to ferro (F) type, and eventually the F-type orbital correlation
causes the spin-singlet formation.
The experimental data are explained by the effective model 
semiquantitatively.

Besides these results, we have also found that there is a chance 
to realize another type of ordered state in the effective model.
Namely, if the Hund's-rule coupling is relatively large, 
the AF orbital correlation may overcome 
the AF spin correlation, yielding the spin-F ordered state 
with AF-type orbital ordering at low temperatures.
This result indicates that the model may exhibit a phase transition 
between the spin-singlet and orbital-F phase 
and the spin-F and orbital-AF phase
by changing the strength of the Hund's-rule coupling.

The aim of this paper is to study 
the phase transition in the effective spin-orbital model in detail. 
Using the MF analysis and the numerical quantum transfer matrix (QTM) method, 
we clarify the nature of the ground-state phase transition and 
the spin-orbital interplay at finite temperatures in the critical regime.
Another purpose of this study is to explore the effect of 
an external magnetic field on the spin-orbital interplay. 
We find that the magnetic field also induces the same-type transition. 

The paper is organized as follows.
In Sec.\ \ref{sec:model}, we introduce the model Hamiltonian. 
The MF results for the ground state are shown 
in Sec.\ \ref{sec:ground-state}.
There, we also discuss a basic mechanism of the phase transitions.
In Sec.\ \ref{sec:numerics}, we present the numerical results 
at finite temperatures obtained by the QTM method.
Section \ref{sec:summary} is devoted to summary and concluding remarks.

\section{Model Hamiltonian}
\label{sec:model}

The microscopic model for the pyroxene compounds has been 
proposed in Ref.\ \citen{Hikihara2004}.
Here, we briefly overview the derivation.

\begin{figure}
\begin{center}
\includegraphics[width=65mm]{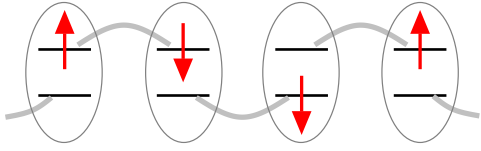}
\end{center}
\caption{
Schematic picture of the 1D multiorbital Hubbard model 
for the pyroxene titanium oxides.
The ellipses represent Ti sites with doubly-degenerate orbitals, 
and the gray curves show the $\sigma$ bonds.
}
\label{fig:model}
\end{figure}

We start from a $t_{2g}$ multiorbital Hubbard model.
In the pyroxene compounds, due to the skew chain structure, 
the TiO$_6$ octahedra are distorted and 
the threefold $t_{2g}$ levels are split into a low-lying doublet 
and a single higher level.
\cite{Ninomiya2003}
It is therefore reasonable to consider the 1D Hubbard model 
with doubly-degenerate orbitals.
For the hopping term, 
considering the edge-sharing network of TiO$_6$ octahedra,
we take account of only the $\sigma$-bond transfer integrals, 
i.e., the transfers between 
the nearest-neighbor (NN) pairs with the same orbitals lying 
in the same plane, and neglect other small matrix elements.
Due to the peculiar zigzag structure of 
the TiO$_6$ chain,\cite{Isobe2002,Hikihara2004}
the NN transfer integrals $t_{i,i+1}^{\alpha\beta}$, 
where $i$ is site index
and $\alpha,\beta=1,2$ are orbital indices,
have an alternating form
\begin{eqnarray}
t^{11}_{i,i+1} = t_\sigma \delta_{i,2n+1},~~~
t^{22}_{i,i+1} = t_\sigma \delta_{i,2n},
\nonumber \\
t^{12}_{i,i+1} = t^{21}_{i,i+1} = 0,
\label{eq:transfer}
\end{eqnarray}
where $n$ is an integer.
The schematic picture of the $\sigma$ bonds is shown in Fig.\ \ref{fig:model}.
As for the onsite interactions, we consider 
the intra- and inter-orbital Coulomb repulsions $U$ and $U'$, and 
the Hund's-rule coupling $J_{\rm H}$, 
which satisfy the relation $U = U' + 2J_{\rm H}$.

From the second order perturbation in the strong-coupling limit 
$t_{\sigma} \ll U, U', J_{\rm H}$, 
the effective spin-orbital Hamiltonian 
(Kugel-Khomskii type\cite{Kugel1982}) is obtained as 
\begin{eqnarray}
\mathcal{H}_{\rm so} &=& -J \sum_{i} \ 
(h_{i,i+1}^{\rm oAF} + h_{i,i+1}^{\rm oF})
- h \sum_i S_i^z, 
\label{eq:H_so}
\\
h_{i,i+1}^{\rm oAF} &=& (A + B \mib{S}_i \cdot \mib{S}_{i+1})
\left(\frac{1}{2} - 2T_i T_{i+1}\right),
\label{eq:h^oAF in T}
\\
h_{i,i+1}^{\rm oF} &=& 
C \left(\frac{1}{4} - \mib{S}_i \cdot \mib{S}_{i+1}\right) 
\nonumber \\
&& \times 
\left[\frac{1}{2} +(-1)^i T_i\right]\left[\frac{1}{2} +(-1)^i T_{i+1}\right],
\label{eq:h^oF in T}
\end{eqnarray}
where 
$\mib{S}_i$ is the $S=1/2$ spin operator 
while $T_i = \pm 1/2$ is the Ising isospin 
which describes two orbital states at site $i$. 
Note that the orbital isospin interactions are of Ising type 
because the transfer integrals (\ref{eq:transfer}) 
are orbital diagonal and 
only one of $t^{11}_{i,i+1}$ or $t^{22}_{i,i+1}$ is nonzero on each bond.
In Eq.~(\ref{eq:H_so}), we included the term for the Zeeman coupling 
with the external magnetic field $h$ for later use. 
The coupling constants in Eqs.~(\ref{eq:H_so})-(\ref{eq:h^oF in T}) 
are given by parameters 
in the original multiorbital Hubbard model as
$J = (t_{\sigma})^2/U$,
$A = 3/4(1-3\eta) + 1/4(1-\eta)$, 
$B = 1/(1-3\eta) - 1/(1-\eta)$, and
$C = 4[1/(1+\eta) + 2/(1-\eta)]/3$ 
where $\eta = J_{\rm H}/U$. 
Hereafter, we set $J = 1$ as an energy unit and 
use the convention of the Boltzmann constant $k_{\rm B}=1$.

We note that 
the JT-type orbital-lattice couplings are also present in real compounds.
For the pyroxene compound NaTiSi$_2$O$_6$, 
the JT stabilization energy is estimated to be $\sim 90$K, which is 
considerably smaller than $J \sim 200$-$300$K.\cite{Hikihara2004}
In this paper, for simplicity, we consider the 1D spin-orbital Hamiltonian 
$\mathcal{H}_{\rm so}$ only.
Indeed, for the realistic value of the JT coupling,
the low-temperature properties are dominated by 
the spin-orbital interplay intrinsic to 
$\mathcal{H}_{\rm so}$, and 
the JT coupling just assists to establish the long-range orderings 
at finite temperatures.\cite{Hikihara2004}

\section{Mean-field analyses of ground-state phase transitions}
\label{sec:ground-state}

\subsection{At zero magnetic field}
\label{sec:MF}
\begin{figure}
\begin{center}
\includegraphics[width=80mm]{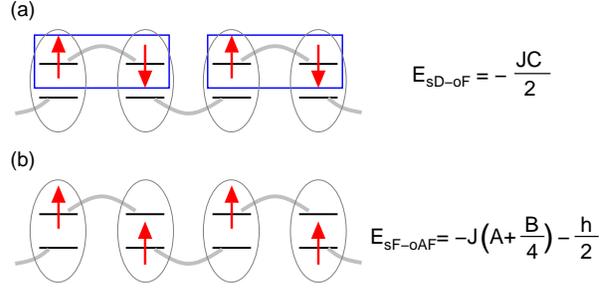}
\end{center}
\caption{
Schematic picture of (a) the sD-oF state and (b) the sF-oAF state.
The rectangles in (a) represent the spin-singlet pairs. 
The ground-state energies per site are also shown.
}
\label{fig:state}
\end{figure}
In this section, we consider the ground-state properties of 
the model (\ref{eq:H_so}) at zero magnetic field $h=0$ 
by using a MF-type argument.
As discussed in the previous study,\cite{Hikihara2004} 
we may expect two different types of long-range ordered states 
in the realistic range of parameters. 
When the orbital ordering is F-type, {\it e.g.}, 
$\langle T_i \rangle = 1/2$ for all $i$, 
Eqs. (\ref{eq:h^oAF in T}) and (\ref{eq:h^oF in T}) show that
the chain is disconnected at every other bonds.
The spin-exchange interactions for remaining isolated NN pairs are 
antiferromagnetic and form the two-site spin-singlet states, 
resulting in the spin-dimer and orbital-F (sD-oF) ground state 
[see Fig.\ \ref{fig:state}(a)].
On the other hand, if the orbital ordering is AF type, {\it e.g.}, 
$\langle T_i \rangle = (-1)^i/2$, 
the system becomes a uniform ferromagnetic spin chain.
Hence, the ground state is the spin-F and orbital-AF (sF-oAF) state 
[Fig.\ \ref{fig:state}(b)].
Comparing the ground-state energies shown in Fig.~\ref{fig:state} at $h=0$, 
we find that the ground state is 
sD-oF for $\eta < \eta_{\rm c}$ 
while sF-oAF for $\eta > \eta_{\rm c}$.
The critical point is $\eta_{\rm c} = (\sqrt{73}-8)/3 \sim 0.181$. 
This transition controlled by $\eta$ is of first order 
and accompanied with finite jumps of the spin gap (from $JC$ to 0) 
and the magnetization (from 0 to the saturated value $\pm 1/2$).

It should be noticed that 
the ground-state energies obtained above fully include quantum fluctuations 
since the orbital interactions are classical Ising type and
since the spin ordering is fully polarized F type (no quantum fluctuation) or 
the two-site singlet state (whose energy is $-JC$ per singlet pair exactly).
In this sense, these results are exact 
unless an ordering with a period longer than two-site spacing 
takes place.
We will comment on the possibility of the ordering 
of longer period in Sec.\ \ref{sec:summary}.

\subsection{At finite magnetic fields}
\label{sec:MF finite h}

The transition between the sD-oF and sF-oAF phases 
is also driven by the external magnetic field $h$. 
The magnetic field suppresses AF spin correlations 
and favors spin-F state,
and hence 
we may expect 
a field-induced transition from the sD-oF phase to the sF-oAF phase. 
Comparing the energies in Fig.~\ref{fig:state} 
including the Zeeman contribution, 
the critical field $h_{\rm c}$ is calculated as 
\begin{equation}
h_{\rm c} = \frac{2(3-16\eta-3\eta^2)}{3(1-3\eta)(1+\eta)(1-\eta)}.
\label{eq:h_c}
\end{equation}
This is also the first-order transition; 
at the critical point, all the isolated singlet pairs collapse at once, 
leading to the fully polarized state.
We note that the result (\ref{eq:h_c}) is also exact in the same sense 
as that in Sec.\ \ref{sec:MF}.
The phase diagram is shown in Fig.~\ref{fig:phasediagram} 
in comparison with numerical results.

\subsection{Competition between AF superexchange interaction and Hund's-rule coupling}
\label{sec:competition}
\begin{figure}
\begin{center}
\includegraphics[width=85mm]{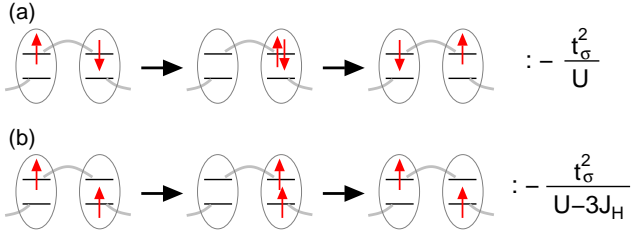}
\end{center}
\caption{
Schematic picture of typical processes of the virtual hoppings 
in (a) the sD-oF state and (b) the sF-oAF state.
The energy gains for each process are also shown.
}
\label{fig:process}
\end{figure}
The phase transitions discussed in the preceding sections 
can be viewed as a result of 
the competition between the spin-superexchange interaction and 
the Hund's-rule coupling. 
The following discussion is sketchy 
but useful to illustrate the competition.  
For simplicity, we consider the case without the magnetic field 
although the discussion is applicable to the case with a finite $h$.
 
Figure\ \ref{fig:process} shows typical contributions 
in the perturbation processes to derive the model (\ref{eq:H_so}):
(a) shows a typical process for the sD-oF configuration, 
which contributes to the AF spin-superexchange interaction.
The energy of the intermediate state is $U$.
On the other hand, (b) shows a typical process 
for the sF-oAF configuration, whose intermediate energy is  
$U'-J_{\rm H} = U-3J_{\rm H}$. 
Hence, the Hund's-rule coupling $J_{\rm H} > 0$ tends to 
enhance the energy gain of the latter process and stabilize the sF-oAF state.
We note that the sF-oAF state is indeed favored 
for any positive Hund's-rule coupling ($\eta = J_{\rm H}/U> 0$) 
in the two-orbital model with uniform transfer integrals, 
i.e., 
$t^{11}_{i,i+1} = t^{22}_{i,i+1} = t_\sigma > 0$ for all NN bonds.
\cite{Roth1966,Inagaki1973} 
On the contrary, in the present system, the energy gain by 
the superexchange process (a) is enhanced in the sD-oF state: 
Since the alternating transfer integrals (\ref{eq:transfer}) 
break up the system into independent two-site spin-singlet pairs, 
the ground-state energy for each connected bond is amplified up to $-JC$. 
Moreover, in the AF orbital configurations (b), 
there is no energy gain from the quantum singlet nature 
because the orbital isospins are of Ising type.
These facts give rise to a finite regime of the sD-oF phase 
for the small values of $\eta$.

\section{Numerical results at finite temperatures} 
\label{sec:numerics}
In this section, we present our numerical results for 
thermodynamic properties of the model (\ref{eq:H_so}) 
obtained by the QTM 
method.\cite{Betsuyaku1984,Betsuyaku1985} 
Details of the method is described in Ref.\ \citen{Hikihara2004}.
In the following, we calculate the NN 
spin and orbital correlations,
$C_{\rm s} = \langle \mib{S}_i \cdot \mib{S}_{i+1} \rangle$ and
$C_{\rm o} = \langle T_i T_{i+1} \rangle$, respectively, 
to clarify the spin-orbital interplay at finite temperatures and 
the instabilities toward the ground states. 
The data are for the Trotter number $M = 4$ 
and we have checked the $M$-convergence of the data.

\subsection{At zero magnetic field}
\label{sec:QTM zero field}

\begin{figure}
\begin{center}
\includegraphics[width=75mm]{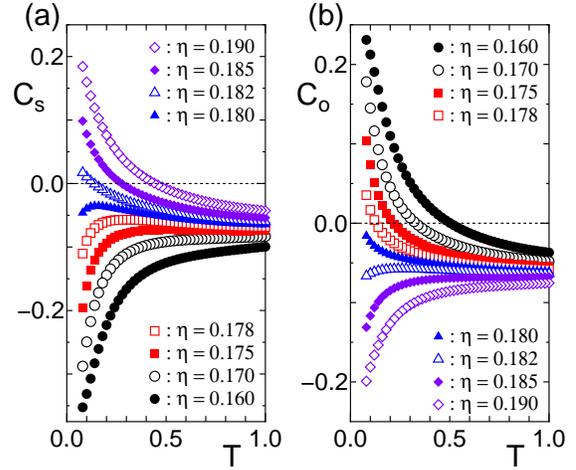}
\end{center}
\caption{
(a) Nearest-neighbor spin correlation $C_{\rm s}$ 
and (b) the nearest-neighbor orbital correlation $C_{\rm o}$
for $0.160 \le \eta \le 0.190$.
}
\label{fig:corzeroH}
\end{figure}
First, we show the numerical results in the absence of the magnetic field.
Figure \ref{fig:corzeroH} shows the data of $C_{\rm s}$ and $C_{\rm o}$ 
for several values of $\eta$.
At high temperatures, both spin and orbital correlations 
are negative, i.e., AF type, and compete with each other.
For $\eta \lsim 0.180$, the AF spin correlation wins the competition 
and yields the sign change of the orbital correlation from AF to F type. 
After the sign change, both $C_{\rm s}$ and $C_{\rm o}$ 
grow cooperatively as temperature decreases, and eventually appear to approach 
$C_{\rm s} = -3/8$ and $C_{\rm o} = 1/4$, 
the values expected for the sD-oF ground state.
On the other hand, for $\eta \gsim 0.182$, the AF orbital correlation 
wins the competition and the spin correlation is changed from AF to F type.
The correlations appear to converge to 
$C_{\rm s} = 1/4$ and $C_{\rm o} = -1/4$ as $T \to 0$,
indicating that the system shows the sF-oAF ground state.
These results suggest that the ground-state phase transition between 
the sD-oF and sF-oAF phases occurs at $\eta = \eta_{\rm c} \simeq 0.181$, 
being consistent with the result of the MF analysis.

It is remarkable that for $\eta$ close to $\eta_{\rm c}$ 
the competition between the orbital and spin correlations 
remains serious down to very low temperatures $T \sim 0.1$; 
both the correlations are gradually {\it suppressed} 
as temperature decreases, and 
the convergence to the values expected for the ordered states 
is not so clear. 
This peculiar behavior of the correlations indicates that 
the keen competition between orbital and spin 
severely affects the finite-temperature properties 
in the multicritical regime $\eta \sim \eta_{\rm c}$.

\subsection{At finite magnetic fields: phase diagram}
\label{sec:field}

\begin{figure}
\begin{center}
\includegraphics[width=75mm]{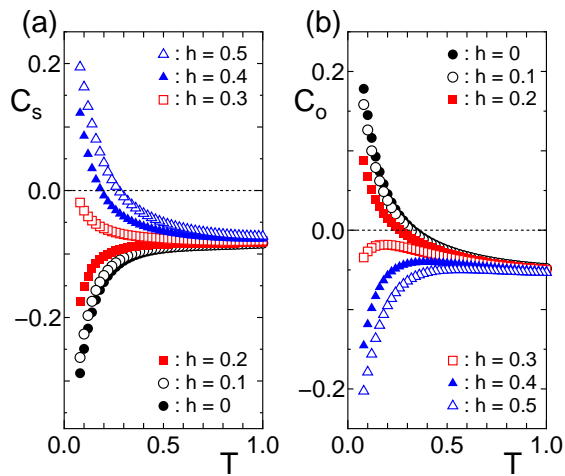}
\end{center}
\caption{
(a) Nearest-neighbor spin correlation $C_{\rm s}$ 
and (b) the nearest-neighbor orbital correlation $C_{\rm o}$
for $\eta = 0.170$ and several typical values of $h$.
}
\label{fig:corfiniteH}
\end{figure}

\begin{figure}
\begin{center}
\includegraphics[width=60mm]{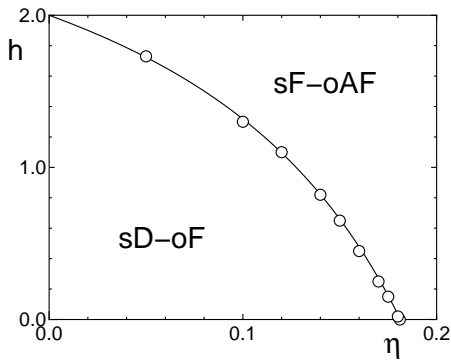}
\end{center}
\caption{
Ground-state phase diagram in the $\eta$ versus $h$ plane.
The solid curve represents the mean-field phase boundary [Eq.\ (\ref{eq:h_c})]. 
The circles correspond to the transition points 
estimated by the numerical calculations.
}
\label{fig:phasediagram}
\end{figure}

In Fig.\ \ref{fig:corfiniteH}, we show the results of 
$C_{\rm s}$ and $C_{\rm o}$ for finite magnetic fields at $\eta = 0.17$.
Similarly to the results in Sec.\ \ref{sec:QTM zero field}, 
there occurs the transition from the sD-oF phase to the sF-oAF phase 
by increasing $h$, 
and near the critical field 
the severe spin-orbital competition survives 
down to very low temperatures 
(see the data for $h = 0.3$).
We estimate the critical field as
$h_{\rm c} \sim 0.25$ for $\eta = 0.17$, 
being consistent with the MF result, 
$h_{\rm c} \simeq 0.27$.

Performing the similar calculations for various values of $\eta$, 
we obtain the ground-state phase diagram in the plane of $\eta$ 
versus $h$ in Fig.\ \ref{fig:phasediagram}. 
The estimates of the critical field by the QTM method agree well with 
the MF prediction.

\section{Summary and concluding remarks}
\label{sec:summary}

We have studied the effective spin-orbital model with 
alternating couplings [Eq.\ (\ref{eq:H_so})], which is derived 
for the pyroxene titanium oxides 
$A$TiSi$_2$O$_6$ ($A$ = Na, Li).
Using the MF-type analysis and the numerical QTM method, 
we have shown that the model exhibits the transition 
from the spin-dimer and orbital-F phase to the spin-F and orbital-AF phase 
as the Hund's-rule coupling parameter $\eta = J_{\rm H}/U$ increases.
The critical value of $\eta$ is estimated to be $\eta_{\rm c} \simeq 0.181$.
Moreover, we have also investigated the effect of the magnetic field $h$
on the phase transition and determined the phase diagram 
in the $\eta$ versus $h$ plane.
The results indicate that the field-induced phase transition
occurs in the system with $\eta < \eta_{\rm c}$.
We have found a severe competition between orbital and spin degrees of freedom 
remaining down to very low temperatures near the phase boundary. 

As mentioned in Sec.\ \ref{sec:ground-state}, the present MF analysis of 
the ground state is exact as far as we consider the orderings 
of two-site period. 
The QTM results, which can include longer-period orderings, 
appear to agree well with the MF estimates of the ground-state phase boundary. 
This agreement suggests that there is no intermediate phase. 
However, as we noted in Sec.~\ref{sec:numerics}, 
in the critical regime the QTM results are not so conclusive 
to determine the ground state 
because of the lack of very low-temperature data; 
there the QTM method suffers from systematic errors 
due to the finite Trotter number, and  
the accuracy is assured 
for $T \gsim 0.07$ in the present study. 
Therefore, it is not ruled out to have novel phases 
in the narrow critical region. 
This interesting issue is left for future study.

It is a challenging problem to observe the field-induced transition 
experimentally.
The critical field $h_{\rm c}$ becomes large 
as $\eta$ deviates from $\eta_{\rm c}$; 
from the estimate of $J = 200$-$300$K,
$h_{\rm c}$ reaches $30$-$50$T at $\eta = 0.175$.
Therefore, to observe the transition in the experimentally accessible range of 
the magnetic field, it is needed to synthesize a compound 
with $\eta$ smaller than but close to $\eta_{\rm c}$. 
Once an appropriate compound is prepared, one may observe 
an abrupt change of the magnetization at $h_{\rm c}$, 
from zero to the saturated value.

\section*{Acknowledgment}
We would like to thank M.\ Isobe and H.\ Seo for fruitful discussions. 
This work is supported by NAREGI.

\end{document}